\documentclass[twocolumn,aps,prl,groupedaddress,showpacs,showkeys]{revtex4}
\usepackage{graphicx}
\usepackage{amsmath}
\usepackage{bm}
\usepackage{color}

\begin{document}
\title{Impact of membrane bistability on dynamical response of neuronal populations}
\author{Wei Wei$^{1,2}$, Fred Wolf$^{3}$ and Xiao-Jing Wang$^{1,2,4}$}
\affiliation{
$^1$ Center for Neural Science, New York University, New York, NY 10003\\
$^2$ Department of Neurobiology and Kavli Institute for Neuroscience, Yale University School of Medicine, New Haven CT 06520\\
$^3$ Max Planck Institute for Dynamics and Self-Organization and Bernstein Center for Computational Neuroscience, D-37077, G\"{o}ttingen, Germany\\
$^4$ NYU-ECNU Institute of Brain and Cognitive Science, NYU Shanghai, Shanghai, China
}

\begin{abstract}
Neurons in many brain areas can develop pronounced depolarized state of membrane potential
(up state) in addition to the normal hyperpolarized down state near the resting potential. The influence of the
up state on signal encoding, however, is not well
investigated. Here we construct a one-dimensional bistable neuron
model and calculate the linear response to noisy oscillatory inputs
analytically. We find that with the appearance of an up state, the transmission function is enhanced by the emergence of a local maximum at some optimal frequency and the phase lag relative to the input signal is reduced. We characterize the dependence of the enhancement of frequency response on intrinsic dynamics and on the occupancy of the up state. 
\end{abstract}

\pacs{87.19.ll, 05.40.-a, 87.19.ls}

\keywords{up and down states, bistability, linear response}

\maketitle

\pagenumbering{arabic} 
\textit{Introduction-} 
Elevated state of neuronal membrane potential (MP), the so called up state, has
been observed extensively in different brain areas
\cite{exp}.  In this regime the MPs of neurons are characterized by a bimodal distribution resulting from two stable fixed points in membrane dynamics. 
This bistability of neuronal dynamics leads to synchronous transitions between the down and up states 
of neurons in a network and the development of global up and down states \cite{parga,compte}. The exact role of the bistability of membrane potential in signal encoding and processing is still not well understood. 

Individual neurons in a network receive noisy synaptic inputs and fire spikes irregularly \cite{shadlen}. Besides the stationary firing rate, 
one important characteristic of neuronal dynamics is the response to time varying signals superimposed on background noise \cite{knight}. This dynamical response of 
cortical neurons has recently been measured experimentally up to 1 kHz of signal frequency, which revealed very hight cutoff frequencies \cite{kondgen}. Theoretically, the linear
response has been obtained analytically for the leaky
integrate-and-fire (LIF) neuron \cite{brunel}, in which
membrane dynamics has only one stable fixed point, and for the
$r-\tau$ model (a piecewise linear version of experiential integrate-and-fire model) \cite{wei}, in which an additional unstable fixed
point for action potential initiation was included. The effect of the unstable fixed point in membrane dynamics on dynamical response  was also investigated numerically in other one-dimensional models and conductance-based models \cite{num}. 
Theoretical characterization of dynamical response of neurons that exhibit up and down
states, however, is still missing. Intuitively, when the membrane potential of a neuron
has a higher probability to be around some depolarized voltage,
it is more likely that a small oscillatory signal can contribute to the firing of spikes, leading to enhancement of frequency response. 
Here we propose an analytically solvable
bistable neuron model to investigate the impact of the up
state on the dynamical response of neurons.

\textit{Model description-} 
In this work we construct a one-dimensional bistable neuron model, which has piecewise linear
subthreshold dynamics and is analytically solvable for the linear
response. The dynamics is described by the following Langevin
equation,
\begin{eqnarray}\label{lge}
\tau\dot{v}&=&f(v)+\mu+\sigma\eta(t)\;,
\end{eqnarray}
where
\begin{eqnarray}\label{lge-fv}
f(v)&=&\left\{ \begin{aligned} &-v\;,\;\qquad -\infty<v \le v_0 \\
&r_1(v-v_{t1})\;,\;v_0<v \le v_1\\
&r(v-v_{t0}) \;,\;\; v_1<v\le v_b
\end{aligned}
\right.
\end{eqnarray}
Here $v$ is the MP relative to the resting potential, $\mu$ is the mean external input, $\eta(t)$ is a Gaussian
white noise which satisfies $\langle\eta(t)\rangle=0$ and
$\langle\eta(t)\eta(t^{'})\rangle=\tau\delta(t-t^{'})$, and $\sigma$ is the strength of the noise. $\tau$ is
the membrane time constant near the resting potential,
which we will take as the unit of time in the theoretical result.
Note that the membrane dynamics here might result from an interaction
between patterned synaptic inputs and intrinsic membrane dynamics.
Fig. 1A shows an illustration of the model dynamics
when there is no external input.  The slopes of the middle and right pieces are denoted as $r_1$ and $r$, respectively. Note that $r$ characterizes the membrane dynamics around the higher stable fixed point where the time constant is given by $\tau/|r|$. MPs at the crossing points of 
the left piece with the middle piece, and the middle piece with the right piece are denoted as $v_0$ and $v_1$, respectively. 
When there is no external input, $v_{t1}$ and $v_{t0}$ are the
unstable fixed point and the higher stable fixed point, respectively. We will fix $v_0$ and $v_{t0}$, whereas
$v_{t1}$ and $v_1$ are given by $v_{t1}=(1+1/r_1)v_0$ and
$v_1=(r_1v_{t1}-rv_{t0})/(r_1-r)$. The deterministic dynamics ($\sigma=0$ in Eq. (\ref{lge})) possess one lower stable fixed point, 
one unstable fixed point and one higher stable fixed point, located at $v=\mu,\; v_{t1}-\mu/r_1\; \textrm{and}\; v_{t0}-\mu/r$, respectively. 
When the MP reaches an absorbing boundary $v_b$, it is reset to a
resetting potential $v_r$ for a refractory period
$\tau_r$. The larger one between $|\tilde{v}_0|$ and $|\tilde{v}_b|$
determines the rheobase current of the model neuron, where
$\tilde{v}_0\equiv-v_0$ and $\tilde{v}_b\equiv r(v_b-v_{t0})$. Since the model neuron will never fire spikes when $|r|$ is very large if $v_b$ is fixed, 
we will fix $\tilde{v}_b$ and choose $v_b$ determined by $v_b=v_{t0}+\tilde{v}_b/r$. The
relative values of $\tilde{v}_0$ and $\tilde{v}_b$ might indicate
different dynamical regime of the model neuron. For example, a larger
$|\tilde{v}_b|$ implies that the neuron can make a transition from the
up-down regime to a tonically depolarized state when
$\mu>|\tilde{v}_0|$, which is reminiscent to that observed for cortical neurons from sleep to wakefulness \cite{steriade}. We are most interested
in the noise-driven regime, i.e., the mean external current is
smaller than the rheobase current, since real neurons work in a regime in which 
excitatory and inhibitory synaptic inputs to individual 
neurons balance each other \cite{somp}. In this dynamic regime, our neuron model
describes barrier penetration in a double well with reinjection of
probability current after reaching an absorbing boundary. Here we
focus on the case when $v_r$ is located within the middle piece, therefore the model neuron could fire several spikes before transiting to
the down state, as normally observed for neurons exhibiting up state. Fig. 1B
shows the MP trajectories for three different $r_1$. We see that with a larger $r_1$, the MP spends more times in
the up state.
\begin{figure}[tbp]
  \centering
  \includegraphics[width=0.45\textwidth]{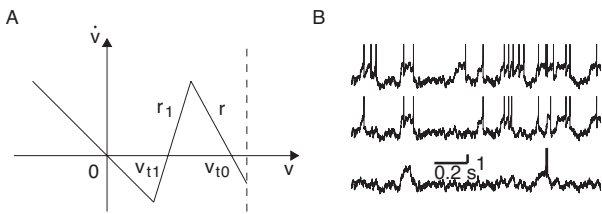}
  \caption{\label{traj-fig} Illustration of the model.
  A, illustration of the piecewise linear model; B, MP trajectories for $r_1=1, 5 \;\textrm{and}\; 10$ from bottom to top.
  Parameters used are:
  $r=-1,\; v_0=0.5,\; v_r=v_{t1},\; v_{t0}=2,\;\tilde{v}_b=-0.2,\; \tau=10\;\textrm{ms},\;\tau_r=0\;\textrm{ms},\;\mu=0,\;\sigma=0.5$.}
\end{figure}

\textit{The Fokker-Planck equation (FPE) framework-} The FPE
corresponding to Eq. (\ref{lge}) has the following form
\cite{risken},
\begin{eqnarray}\label{fpe}
\partial_t{P}(v,t)
+\partial_v(f(v)+\mu-D\partial_v)P(v,t)=0\;,
\end{eqnarray}
where $D=\frac{1}{2}\sigma^2$ is the diffusion constant. We will use
both $D$ and $\sigma$ in the following.
Defining the probability current as
$J(v,t)=(f(v)+\mu-D\partial_v)P(v,t)$,
the FPE then becomes the equation for probability conservation,
$\partial_tP(v,t)+\partial_v J(v,t)=0$.

The boundary conditions are specified in the following (subscripts
1, 2, 3 indicate the left, right, and middle MP
regions in Fig. 1A). At the absorbing boundary $v_b$,
$P_2(v_b,t)=0$. 
At the resetting point $v_r$, $P_3(v_r^+,t)-P_3(v_r^-,t)=0$ and  
$\partial_v P_3(v_r^+,t)-\partial_v P_3(v_r^-,t)=\partial_v
P_2(v_b,t-\tau_r)$, from the the resetting condition and continuity of the probability density and probability
current. At $v_0$ and $v_1$,
the probability density and its derivative are continuous:
$P_1(v_0,t)=P_3(v_0,t)$, $\partial_v P_1(v_0,t)=\partial_v P_3(v_0,t)$,
$P_3(v_1,t)=P_2(v_1,t)$, and $\partial_v P_3(v_1,t)=\partial_v P_2(v_1,t)$.
Finally the normalization condition of the probability density
requires
$\lim_{v\to -\infty}P_1(v,t)=0$.
With these boundary conditions the asymptotic solution of the FPE is uniquely determined (the possible transient is not
of interest here). The instantaneous firing rate is
given by the probability current through the absorbing boundary, 
$\nu(t)\equiv J(v_b,t)=-D\partial_v P_2(v_b,t)$.

\textit{Development of the up state-} When the mean input to a model neuron is
constant, the stationary probability density, denoted as $P_0(v)$,
can be obtained by setting $J(v,t)=\nu_0$ \cite{sup}.
Here $\nu_0$ is the stationary firing rate, which is determined
by the normalization condition of the stationary density,
$\int_{-\infty}^{v_b} P_0(v)\;dv=1$. 
The existence of the up state requires the
appearance of a local maximum of the probability density at a
depolarized MP value.  
Two peaks appear in $P_0(v)$ if there exists an up state in the MP trajectories, located around the lower stable fixed point and the higher stable fixed point, which will be denoted as $P_0^{down}$ and $P_0^{up}$, respectively. The MPs corresponding to the two peaks, denoted as $v_{down}$ and $v_{up}$, are the mean values of MPs at the down state and up state, respectively. From the expression of $P_0(v)$ \cite{sup}, it is easily to see that the down state locates at the lower stable fixed point, $v_{down}=\mu$.  
The development of a local maximum around the higher stable fixed point requires $P_{02}^{'}(v)=0$ since the probability density decreases
monotonically when $0<v<v_1$, or
equivalently, the following equation
\begin{eqnarray}\label{vup}
xe^{-x^2}\int_{x_b}^x\;e^{{x^{'}}^2}\;dx^{'}=\frac{1}{2}
\end{eqnarray}
has a solution within the range $v_1< v<v_b$, where $x\equiv\frac{r(v-v_{t0})+\mu}{\sqrt{-r}\sigma}$, $x_b\equiv\frac{r(v_b-v_{t0})+\mu}{\sqrt{-r}\sigma}$, with $v_{up}$ independent of $r_1$.  Since $x_b<x$, we have $x>0$ and $v_1< v_{up}<v_{t0}-\mu/r$ from Eq. (\ref{vup}). 
Therefore the mean value of the MP at the up state locates lower than the higher stable fixed point in the deterministic dynamics due to the influence of noise and the absorbing boundary.  The probability density at $v=v_{up}$ is given by 
\begin{eqnarray}\label{pup}
P_0^{up}=\frac{\nu_0}{r(v_{up}-v_{t0})+\mu}\;.
\end{eqnarray}

The dependence of $P_0(v)$ on $r_1$ and $r$ is shown in Fig. 2A-B. With the increase of $r_1$, the transition from the up state to the down state becomes more difficult, therefore the MP resides on the up state for a longer time (Fig. \ref{traj-fig}B), indicating a larger ratio between the maximal probability density at the up state and down state (Fig. 2A).  Note that changing the slope $r_1$ can determine whether the up state exists or not by
adjusting $v_1$, but does not influence the position of $v_{up}$ if it exists. With the increase of $|r|$ (the absolute value of $r$), $v_{up}$ is shifted slightly towards the higher deterministic fixed point and the ratio $P_0^{up}/P_0^{down}$ decreases (Fig. 2B).

\textit{Linear response-} 
Now consider a weak sinusoidal signal
encoded in the mean input, $\mu(t)=\mu+\varepsilon \cos(\omega t)$,
where $\varepsilon$ is small. At the linear order in $\varepsilon$,
the instantaneous firing rate is given by
$\nu(t)=\nu_0+\varepsilon |\nu_{1c}(\omega)|\cos(\omega t
-\phi_c(\omega))$,
where $|\nu_{1c}(\omega)|$ is the transmission function and
$\phi_c(\omega)$ is the phase lag. We find that a complex response
function $\nu_{1c}(\omega)$ can be obtained analytically by solving
the FPE at the linear order using the Green's function method. The
transmission function is the absolute value of $\nu_{1c}(\omega)$,
while the phase lag $\phi_c(\omega)$ is given by the phase angle,
$\phi_c(\omega)=\arg(\nu_{1c}(\omega))$. The expression of
$\nu_{1c}(\omega)$ reads
\begin{eqnarray}\label{nu1c}
&&\nu_{1c}(\omega)=\frac{1}{B}\bigg[\frac{i\omega(1+1/r_1)}{(1-i\omega)(1+i\omega/r_1)}
(\psi_1P_{01}-\sqrt{D}\Phi_1P_{01}^{'})\nonumber\\
&&+\frac{i\omega(1/r_1-1/r)}{(1+i\omega_/r)(1+i\omega/r_1)}(\psi_1
Y_5^{'}-\psi_1^{'}Y_5)P_{02}(v_1)\;e^{\Delta_1}\nonumber\\
&&+\frac{\sqrt{Dr_1}(r-r_1)}{(r_1+i\omega)(r+i\omega)}(\psi_1
Y_{51}^{'}-\psi_1^{'}Y_{51})P_{02}^{'}(v_1)\;e^{\Delta_1}\nonumber\\
&&+
\frac{\nu_0/\sqrt{Dr_1}}{1+i\omega/r_1}(\psi_1Y_{51r}^{'}-\psi_1^{'}Y_{51r})e^{\Delta_0}
-\frac{\nu_0/\sqrt{D|r|}}{1+i\omega/r}\times\nonumber\\
&&\quad((\psi_1 Y_5^{'}-\psi_1^{'}Y_5)Y_{2}^{'}-(\psi_1
Y_6^{'}-\psi_1^{'}Y_6)Y_{2})e^{\Delta_2}\bigg]\;,
\end{eqnarray}
where $\psi_{1}(v)$, $\Phi_1(v)$, etc. are parabolic cylinder
functions \cite{abramowitz}, and $Y_1(v)$, $Y_5(v)$, $B$, etc. are
combinations of them to simplify the expressions, as defined in
\cite{sup}. Note the functions adopt their values at $v=v_0$, unless
denoted otherwise. Taking $\omega\to\infty$ in Eq. (\ref{nu1c}), we find that the high frequency limit is the same as the LIF model and the $r-\tau$ model, i.e., $\nu_{1c}\to \frac{\nu_0}{\sqrt{D}}\frac{1}{\sqrt{\omega}}\;e^{i\frac{\pi}{4}}$. This high frequency limit is characteristic of the linear dynamics and absorbing boundary \cite{wei}.
\begin{figure}[tbp]
  \centering
  \includegraphics[width=0.45\textwidth]{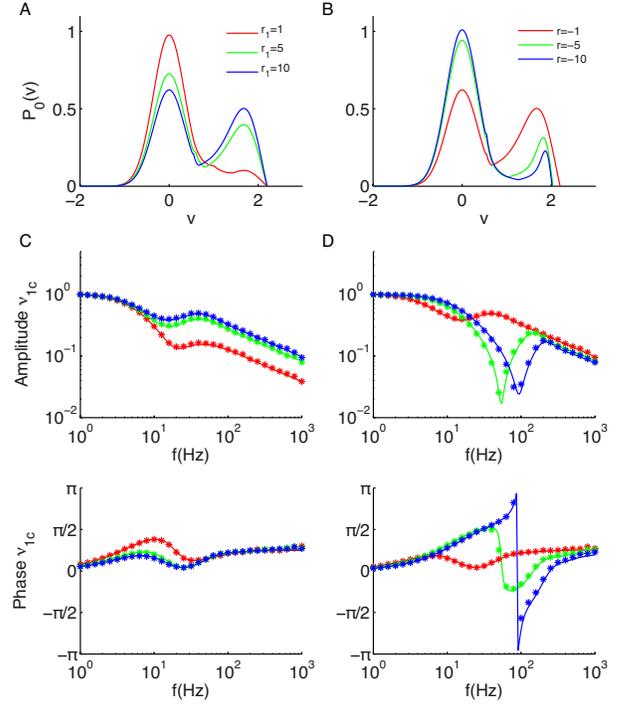}
  \caption{\label{fig2} Dependence of probability density and linear response on $r_1$ and $r$. A and B, probability density for different $r_1$ and $r$. C and D, dependence of the transmission function (upper panels) and phase lag (lower panels) of the linear response on $r_1$ and $r$.  The linear response is normalized with the value at $f=1\;\textrm{Hz}$.  Signal frequency $f$ is related to the angular frequency $\omega$ by $\omega=2\pi f$. Solid lines are from theoretical results and asterisks are from simulations. Parameters used: $r=-1$ in A and C, $r_1=10$ in B and D. Other parameters are the same as in Fig. \ref{traj-fig}. }
\end{figure}

\textit{Frequency-selective enhancement of linear response by the up state-} 
The dependence of the linear response on $r_1$ and $r$ are shown in Fig. 2C-D. We see that a local maximum, denoted as $\nu_{1c}^{max}$, appears at frequency $f^{max}$  (Fig. 2C-D, upper panels) accompanying the development of the up state (Fig. 2A-B).  
The local maximal value of transmission function at the resonance frequency increases with $r_1$ and decreases with $|r|$, following the same trend as the ratio between the probability density at $v_{up}$ and $v_{down}$ (Fig. 2A-B). The phase lag of the firing rate response relative to the input signal is reduced when there is a more pronounced up state (Fig. 2C-D, lower panels). Therefore the up state can enhance the dynamical response by developing local maximum at some specific resonance frequency, and reduce the phase lag of the response. 

We characterize the relationship between the up state occupancy and resonance in the linear response quantitatively in Fig. 3. While the ratio between probability densities at $v_{up}$ and $v_{down}$ increases with $r_1$ (Fig. 3A), the maximal value of the transmission function at the resonance frequency also increases with $r_1$ (Fig. 3C). This leads to an increase of $\nu_{1c}^{max}$ with $P_0^{up}/P_0^{down}$ (Fig. 3C, inset). Similarly, $\nu_{1c}^{max}$ decreases with $|r|$ (Fig. 3D), following with the same trend as $P_0^{up}/P_0^{down}$ except for an initial small $|r|$ regime where there is no local maximum for $|\nu_{1c}(\omega)|$ (Fig. 3B). This also leads to an increase of $\nu_{1c}^{max}$ with $P_0^{up}/P_0^{down}$ (Fig. 3D, inset). The signal frequency at which the transmission function is maximally enhanced, $f^{max}$, keeps constant when $r_1$ increases, i.e., being independent of the time scale characterizing the unstable fixed point. On the contrary, $f^{max}$ goes to higher frequencies when $|r|$ increases, therefore is determined by the membrane time constant at the up state, $\tau/|r|$.
  
\begin{figure}[tbp]
  \centering
  \includegraphics[width=0.45\textwidth]{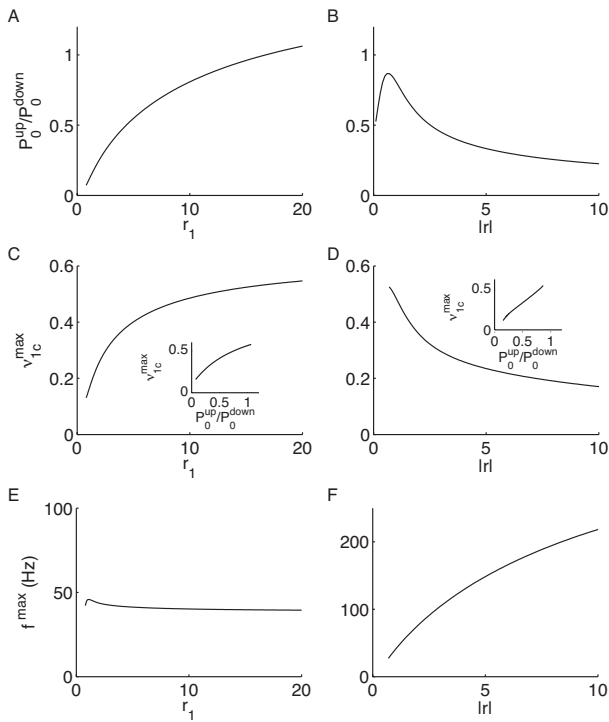}
  \caption{\label{nu1r1-fig} Dependence of up state occupancy and resonance in the linear response on $r_1$ and $r$. A and B, the up state Occupancy $P_0^{up}/P_0^{down}$ as a function of $r_1$ (A) and $|r|$ (B). C and D, the local maximal value $\nu_{1c}^{max}$ of the normalized transmission function as a function of $r_1$ (C) and $|r|$ (D). E and F, the resonance frequency $f^{max}$ as a function of $r_1$ (E) and $|r|$ (F). Insets: $\nu_{1c}^{max}$ as a function of $P_0^{up}/P_0^{down}$ due to the change of $r_1$ (C) and and $|r|$ (D). Parameters used: $r=-1$ in A and $r_1=10$ in B. Other parameters are the same as in Fig. \ref{traj-fig}.  }
\end{figure}

We tested the predication by using a neuron model with biophysically
realistic membrane currents, which includes a noninactivating
potassium current controlling  the up state and an inward rectifying
potassium current stabilizing the down state \cite{compte}. We find
that the normalized frequency response is
strongly enhanced with the development of the up state \cite{sup}.

\textit{Discussion-} Dynamical response to noisy oscillatory inputs is a fundamental
characterization of neurons' capability in signal encoding and transmission. Our model predicts  a frequency-selective enhancement of signal transmission with the development of the up state.  This prediction can be experimentally tested for neurons exhibiting up and down states by performing experiments similar to that in \cite{kondgen}.
Selective enhancement of specific high frequency component of signals might be functionally beneficial for neural communication, since signals with frequencies within the gamma band (30-100 Hz) or ``high gamma" ($>100$ Hz) have been suggested to synchronize inter-regional brain activity \cite{gamma}. 
Bistable piecewise linear membrane dynamics was
introduced previously to approximate the nullcline of MP in the FitzHugh-Nagumo
model \cite{mckean}. In the limit of $\tau\to 0$, the linear
response of that system was obtained \cite{lindner1}. Here we obtain
the linear response analytically for a one-dimensional bistable system with general $\tau$. 
Bistability in individual neurons provides an alternative explanation
for persistent activity in some brain areas \cite{marder}, and is related to perfect temporal information accumulation \cite{okamoto} and self-organized criticality \cite{millman}.
Our work sheds insights into the possible role of the up state on signal encoding and transmission through
population response. Further work on building a network of interconnected such bistable units is needed to examine the relationship between the up state at the individual neuron level and at the circuit level \cite{stern}.

We thank J. Mejias, J. Murray and F. Song for carefully reading the manuscript.

\end{document}